\documentstyle[12pt]{article}
\begin{document}
\begin{titlepage}

\begin{flushright} 
IHEP-TH-98-02\\
AMES-HET-98-04\\
TU-546 \\
hep-ph/9804343
\end{flushright}
\vspace{0.2in}

\begin{center}
{\Large{ Probing Anomalous Top Quark Couplings At $e\gamma$ Colliders}}
\end{center}
\vspace{.3in}

\begin{center}
Jun-Jie Cao$^{1,2}$, Jian-Xiong Wang$^{1}$, Jin Min Yang$^{3}$, 
Bing-Lin Young$^4$ and Xinmin Zhang$^{5,1}$
\end{center}
         
\vspace{.2in}

\begin{center}
\it
$^1$Institute of High Energy Physics, Academy Sinica,\\
    Beijing 100039, China\\
$^2$Department of Physics, Henan Normal University, \\
    Xinxiang, Henan 453002, China\\
$^3$ Department of Physics, Tohoku University, Sendai 980-8578, Japan\\
$^4$ Department of Physics and Astronomy, Iowa State University, \\
     Ames, Iowa 50011, USA \\
$^5$ CCAST (World Laboratory), P.O.Box 8730, Beijing 100080, China
\end{center}
\vspace{.7in}
\rm
 
\begin{center} ABSTRACT \end{center}

We study in the effective Lagrangian approach the possibility of probing 
anomalous top quark charged current couplings in the single top production 
at a high energy $e\gamma$ collider.  We analyzed all possible dimension-six 
CP-conserving operators which can give rise to an anomalous $Wtb$ coupling
which represent new physics effect generated at a higher energy scale. For 
those operators which also give rise to anomalous $Zb\bar b$ or right-handed 
$Wtb$ couplings, we find that they are strongly constrained by the existing
$R_b$ and $b\rightarrow s\gamma$ data.  As a result, a collider with a
luminosity of the order of 100 $fb^{-1}$ is required to 
observe the anomalous effects.  For other operators which currently subject 
to no strong constraints, the high energy $e\gamma$ collider can probe their 
couplings effectively because of the clean environment of such a collider.

\vfill
PACS: 14.65.Ha, 12.60.Cn
\end{titlepage}
\eject

\baselineskip=0.30in
\begin{center}{\Large 1. Introduction}\end{center}

The standard model (SM) has been proved to be phenomenologically successful.
However, it is also generally believed that the SM will be augmented by new 
physics at higher energy scales.  It is a challenge to search for ways to
reveal the new physics effects at either the existing accelerators or the
possible future ones.  The top quark, because of its large mass, is believed 
to be more sensitive to new physics than other particles [1][2]. Hence, if 
anomalous top quark couplings exist, they may readily manifest themselves in 
the processes of top quark production and decay, and they may also affect
the decay width of Z to $b \bar b$ and b decays. 

To probe the anomalous  $Wtb$ coupling, the single top quark production  
$e\gamma \rightarrow \bar{t}b\nu$ at a high energy $e\gamma$ collider has 
been analyzed in [3][4]. It was argued that because of the low background 
events, the top quark coupling $Wtb$ can be measured with high precision, 
much better than in hadron colliders, the Tevatron and LHC.  Therefore,
more sensitive tests of the $Wtb$ can be achieved in an $e\gamma$ collider.
In these early works, the probing of the $Wtb$ coupling was done in 
isolation,  its possible relations to other couplings and, therefore, 
possible constraints were not taken into account.

In this paper we re-examine the possibility of probing anomalous $Wtb$ at 
an $ e \gamma$ collider in a more structured theoretical framework in which 
other couplings will come into play.  We pay special attention to the effects
of the latter and analyze the experimental constraints from $R_b$ at LEP I
and the experimental data on $b \rightarrow s + \gamma$ at CESR.
Presently the experimental data on $R_b$ is consistent with the prediction 
of the standard model within 1.4$\sigma$[5], which may in turn put strong 
bounds on some of the anomalous couplings of the top quarks.

A systematic, model independent framework for exploring anomalous top quark 
couplings is the effective Lagrangian.  It introduces no new particles and
can be made to deviate from the SM very slightly as required by the current 
data. 
There are two common approaches to the effective Lagrangian of the top quarks
in the literature. They are formulated in terms of either the non-linear or
linear realization of the electroweak symmetry. In the case of linear 
realization, the new physics is parameterized by 
higher dimension operators which contain the fields of the SM and are 
invariant under the SM symmetry, SM $SU_c(3)\times SU_L(2)\times U_Y(1)$.
Above the electroweak symmetry breaking scale but below the new physics scale,
the effective Lagrangian can be written as
\begin{equation}
{\cal L}_{eff}={\cal L}_0+\frac{1}{\Lambda^2}\sum_i C_i O_i
                         +O(\frac{1}{\Lambda^4})
\end{equation}
where ${\cal L}_0$ is the SM Lagrangian, $\Lambda$ is the new physics scale, 
$O_i$ are dimension-six operators which are 
$SU_c(3)\times SU_L(2)\times U_Y(1)$ invariant and $C_i$ are constants which 
represent the coupling strengths of $O_i$.  This expansion was first discussed 
in Ref.[6]. Recently the effective operators involving the top quark were 
reclassified and some are analyzed [7][8].  In this paper, we use such 
linearly realized effective Lagrangian and analyze all possible dimension-six 
CP-conserving operators containing anomalous $Wtb$ couplings. For those 
operators which also give rise to anomalous $Zb\bar b$ or right-handed $Wtb$ 
couplings, we will examine their experimental constraints from the recent 
$R_b$  data at LEP I and from the $b\rightarrow s\gamma$ data at CLEO.  Our 
analyses show that the existing experimental data strongly constrains the 
coupling strength of these operators.
For other 
operators which subject to no strong experimental constraints so far,  we 
find that the high energy $e\gamma$ 
collider can provide an effective probe to their couplings. 

This paper is organized as follows. In Sec. 2, we list the dimension-six
operators which contribute to $Wtb$ couplings.  In Sec. 3 we derive the bounds 
from the data of $R_b$ and $b\rightarrow s\gamma$ for those which give rise 
to anomalous $Zb\bar b$ coupling or right-handed $Wtb$ coupling.  In Sec. 4 
we determine the possibility of probing anomalous $Wtb$ coupling in the 
single top production at the high energy $e\gamma$ collider. Finally, in Sec.5 
we conclude our paper with discussions and a summary.
\vspace{1cm}

\begin{center} {\Large 2. Operators contributing 
                          to $Wtb$ couplings }
\end{center}
The dimension-six operators which contribute to the $Wtb$ couplings are give by
\begin{eqnarray}
O_{qW}&=&\left [\bar q_L \gamma^{\mu}\tau^I D^{\nu}q_L
         +\overline{D^{\nu}q_L} \gamma^{\mu}\tau^I q_L\right ]
          W^I_{\mu\nu},\\
O_{\Phi q}^{(3)}&=&i\left [\Phi^{\dagger}\tau^I D_{\mu}\Phi
        -(D_{\mu}\Phi)^{\dagger}\tau^I\Phi\right ]\bar q_L \gamma^{\mu}\tau^I 
       q_L,\\
O_{Db}&=&(\bar q_L D_{\mu} b_R) D^{\mu}\Phi
         +(D^{\mu}\Phi)^{\dagger}(\overline{D_{\mu}b_R}q_L),\\
O_{bW\Phi}&=&\left [(\bar q_L \sigma^{\mu\nu}\tau^I b_R) \Phi
         +\Phi^{\dagger}(\bar b_R \sigma^{\mu\nu}\tau^I q_L)\right ]
          W^I, \\
O_{t3}&=&i\left [(\widetilde\Phi^{\dagger}D_{\mu}\Phi)(\bar t_R \gamma^{\mu}b_R
         -(D_{\mu}\Phi)^{\dagger}\widetilde\Phi(\bar b_R \gamma^{\mu}t_R)
       \right ],\\
O_{Dt}&=&(\bar q_L D_{\mu} t_R) D^{\mu}\widetilde\Phi
         +(D^{\mu}\widetilde\Phi)^{\dagger}(\overline{D_{\mu}t_R}q_L),\\
O_{tW\Phi}&=&\left [(\bar q_L \sigma^{\mu\nu}\tau^I t_R) \widetilde\Phi
         +\widetilde\Phi^{\dagger}(\bar t_R \sigma^{\mu\nu}\tau^I q_L)\right ]
          W^I_{\mu\nu},
\end{eqnarray}
where we follow the standard notation [7][8]: $q_L$ denotes the third family 
left-handed doublet quarks, $\Phi$ and $\tilde{\Phi}$ are the Higgs field 
and its equivalent complex conjugate representation, $W_{\mu\nu}$ 
and $B_{\mu\nu}$ are the SU(2) and U(1) gauge boson field tensors 
in the appropriate matrix forms, and $D_\mu$ denotes the appropriate 
covariant derivatives.   

The expressions of these operators in the unitary gauge after electroweak 
symmetry breaking are given as 
\begin{eqnarray}
O_{qW}&=&\frac{1}{2}W^3_{\mu\nu}\left [\bar t_L\gamma^{\mu}\partial^{\nu}t_L
+\partial^{\nu}\bar t_L\gamma^{\mu}t_L
-\bar b_L\gamma^{\mu}\partial^{\nu}b_L
-\partial^{\nu}\bar b_L\gamma^{\mu}b_L\right ]  \nonumber\\
& & +\frac{1}{\sqrt 2}\left [W^+_{\mu\nu}(\bar t_L\gamma^{\mu}\partial^{\nu}b_L
+\partial^{\nu}\bar t_L\gamma^{\mu}b_L)
+W^-_{\mu\nu}(\bar b_L\gamma^{\mu}\partial^{\nu}t_L
+\partial^{\nu}\bar b_L\gamma^{\mu}t_L) \right ]\nonumber\\
& & -ig_2\bar q_L \gamma^{\mu} \left [W_{\mu},W_{\nu}\right ]\partial^{\nu}q_L
  -ig_2\partial^{\nu}\bar q_L \gamma^{\mu} 
  \left [W_{\mu},W_{\nu}\right ]q_L
  -ig_2 \bar q_L \gamma^{\mu} \left [W_{\mu\nu},W^{\nu}\right ]q_L,\\
O_{\Phi q}^{(3)}&=&-\frac{1}{2}g_Z(H+v)^2Z_{\mu}\left [\bar t_L\gamma^{\mu}t_L
          -\bar b_L\gamma^{\mu}b_L\right ]\nonumber\\
& &    +\frac {1}{\sqrt 2}g_2(H+v)^2\left [W^+_{\mu}\bar t_L\gamma^{\mu}b_L
                +W^-_{\mu}\bar b_L\gamma^{\mu}t_L\right ],\\
O_{Db}&=&\frac{1}{2\sqrt 2}\partial^{\mu}H \left [\partial_{\mu}(\bar b b)
  +\bar b\gamma_5\partial_{\mu}b-(\partial_{\mu}\bar b)\gamma_5 b
   +\frac{2}{3}g_1B_{\mu}\bar bi\gamma_5  b\right ]\nonumber\\
& & +\frac{i}{4\sqrt 2}g_Z (H+v)Z^{\mu}\left [(\partial_{\mu}\bar b) b
     -\bar b\partial_{\mu}b-\partial_{\mu}(\bar b \gamma_5b)
-i\frac{2}{3}g_1B_{\mu}(\bar b b)\right ]\nonumber\\
& & -\frac{i}{2}g_2 (H+v)\left [W_{\mu}^+ (\bar t_L \partial^{\mu}b_R
                 +i\frac{g_1}{3}B^{\mu} \bar t_L  b_R)
-W_{\mu}^- ( \partial^{\mu}\bar b_R t_L
-i\frac{g_1}{3}B^{\mu} \bar b_R t_L)\right ],\\
O_{bW\Phi}&=&\frac{1}{2}(H+v)\left [W^+_{\mu\nu}(\bar t_L \sigma^{\mu\nu} b_R)
        +W^-_{\mu\nu}(\bar b_R \sigma^{\mu\nu} t_L)
-\frac{1}{\sqrt 2}W^3_{\mu\nu}(\bar b \sigma^{\mu\nu} b)\right.\nonumber\\
& & +ig_2(W^+_{\mu}W^3_{\nu}-W^3_{\mu}W^+_{\nu})(\bar t_L \sigma^{\mu\nu} b_R)
    -ig_2(W^-_{\mu}W^3_{\nu}-W^3_{\mu}W^-_{\nu})(\bar b_R \sigma^{\mu\nu} t_L)
\nonumber\\
& &\left.+i\frac{g_2}{\sqrt 2}
(W^+_{\mu}W^-_{\nu}-W^-_{\mu}W^+_{\nu})(\bar b \sigma^{\mu\nu}b)\right ],\\
O_{t3}&=&\frac{1}{2\sqrt 2}g_2 (H+v)^2\left [
    W_{\mu}^+ (\bar t_R\gamma^{\mu} b_R)
    +W_{\mu}^- (\bar b_R\gamma^{\mu} t_R)\right ],\\
O_{Dt}&=&\frac{1}{2\sqrt 2}\partial^{\mu}H \left [\partial_{\mu}(\bar t t)
    +\bar t\gamma_5\partial_{\mu}t-(\partial_{\mu}\bar t)\gamma_5 t
    -i\frac{4}{3}g_1 B_{\mu}\bar t \gamma_5 t\right ]\nonumber\\
& & +i\frac{1}{4\sqrt 2}g_Z (H+v)Z^{\mu}\left [\bar t\partial_{\mu}t
    -(\partial_{\mu}\bar t) t+\partial_{\mu}(\bar t\gamma_5 t)
-i\frac{4}{3}g_1 B_{\mu}\bar t t\right ]\nonumber\\
& & -i\frac{1}{2}g_2 (H+v)W_{\mu}^- \left [\bar b_L\partial^{\mu} t_R
                 -i\frac{2}{3}g_1 B^{\mu}\bar b_L  t_R\right ]\nonumber\\
& & +i\frac{1}{2}g_2 (H+v)W_{\mu}^+ \left [(\partial^{\mu} \bar t_R)b_L
                 +i\frac{2}{3}g_1 B^{\mu}\bar t_R  b_L\right ],\\
O_{tW\Phi}&=&\frac{1}{2\sqrt 2}(H+v)(\bar t\sigma^{\mu\nu}t)
\left [W^3_{\mu\nu}-ig_2(W^+_{\mu}W^-_{\nu}-W^-_{\mu}W^+_{\nu})\right ]
\nonumber\\
& & +\frac{1}{2}(H+v)(\bar b_L\sigma^{\mu\nu} t_R)
\left [W^-_{\mu\nu}-ig_2(W^-_{\mu}W^3_{\nu}-W^3_{\mu}W^-_{\nu})\right ]
\nonumber \\
& & +\frac{1}{2}(H+v)(\bar t_R\sigma^{\mu\nu} b_L)
\left [W^+_{\mu\nu}-ig_2(W^3_{\mu}W^+_{\nu}-W^+_{\mu}W^3_{\nu})\right ],
\end{eqnarray}
where $g_Z=2m_Z/v=\sqrt {g_1^2+g_2^2}$ with $v$ being the vacuum expectation 
value of the Higgs boson.  

The first four operators induce anomalous $Zb\bar b$ couplings which will 
affect $R_b$.  All operators contain an anomalous $W t\bar b$ coupling and 
the operator in Eq. (15) contains the $W t\bar b \gamma$ coupling.  
Both couplings
contribute to the single top production at the $e\gamma$ collider.
The induced effective vertices for the couplings $Zb\bar b$, $W t\bar b$ 
and $W t\bar b \gamma$ relevant to our analyses are given by
\begin{eqnarray}
{\cal L}_{Zb\bar b}&=&
\frac{C_{qW}}{\Lambda^2}\frac{c_W}{2}
   Z_{\mu\nu}(\bar b \gamma^{\mu}P_L \partial^{\nu}b
                      +\partial^{\nu}\bar b \gamma^{\mu}P_L b)
 +\frac{C_{\Phi q}^{(3)}}{\Lambda^2} (v m_Z) 
    Z_{\mu}(\bar b\gamma^{\mu}P_L b)
\nonumber\\
& &+\frac{C_{Db}}{\Lambda^2}
 \frac{m_Z}{2\sqrt 2}  
    Z^{\mu}\left [i(\partial_{\mu}\bar b b-\bar b\partial_{\mu}b)
                    -i\partial_{\mu}(\bar b \gamma_5b)\right ]
+\frac{C_{bW\Phi}}{\Lambda^2}\frac{c_W}{2} \frac{v}{\sqrt 2} 
   Z_{\mu\nu}(\bar b \sigma^{\mu\nu} b),\\ 
\label{wtb}
{\cal L}_{Wt\bar b}&=&\frac{C_{qW}}{\Lambda^2} \frac{1}{\sqrt 2}
        W^+_{\mu\nu}(\bar t\gamma^{\mu}P_L\partial^{\nu}b
                     +\partial^{\nu}\bar t\gamma^{\mu}P_L b)
+\frac{C_{\Phi q}^{(3)}}{\Lambda^2}\frac {g_2}{\sqrt 2}v^2
      W^+_{\mu}(\bar t\gamma^{\mu}P_L b)\nonumber\\
& &-\frac{C_{Db}}{\Lambda^2}\frac{v}{\sqrt 2}\frac{g_2}{\sqrt 2} 
      W_{\mu}^+ (i\bar t P_R \partial^{\mu}b)
   +\frac{C_{bW\Phi}}{\Lambda^2}\frac{v}{2} 
      W^+_{\mu\nu}(\bar t \sigma^{\mu\nu}P_R b)\nonumber\\
& & +\frac{C_{t3}}{\Lambda^2} \frac{v^2}{2}\frac{g_2}{\sqrt 2} 
               W_{\mu}^+ (\bar t\gamma^{\mu}P_R b)
+\frac{C_{Dt}}{\Lambda^2} \frac{v}{\sqrt 2}\frac{g_2}{\sqrt 2}
     W_{\mu}^+ (i\partial^{\mu} \bar t)P_L b\nonumber\\
& & +\frac{C_{tW\Phi}}{\Lambda^2}\frac{v}{2}
        W^+_{\mu\nu}(\bar t\sigma^{\mu\nu}P_L b),\\
\label{wtbg}
{\cal L}_{Wt\bar b\gamma}&=&\frac{C_{bW\Phi}}{\Lambda^2}
 (i2g_2s_W) W^+_{\mu}A_{\nu}(\bar t \sigma^{\mu\nu}P_R b),
\end{eqnarray}
where  $s_W\equiv \sin\theta_W$ and $c_W\equiv \cos\theta_W$. 
\vspace{1cm}

\begin{center} {\Large 3. Current bounds from experimental data}
\end{center}

For the on-shell $Z$, we obtain the effective vertex $Zb\bar b$ 
\begin{equation}\label{zbb}
 \Gamma_{\mu}=-i\frac{e}{4s_Wc_W} \left [\gamma_{\mu}V
-\gamma_{\mu}\gamma_5 A+\frac{1}{2m_b}(p_b-p_{\bar b})_{\mu}
S\right ],
\end{equation}
where $p_b$ and $p_{\bar b}$ are the
momenta of outgoing quark and anti-quark, respectively. 
We write the vector and axial-vector couplings  as
\begin{eqnarray}
V&=&v_b+\delta V,\\
A&=&a_b+\delta A,
\end{eqnarray}
where $v_b$ and $a_b$ represent the SM couplings
and $\delta V,\delta A$ the new physics
contributions. The SM couplings are given by 
\begin{eqnarray}
v_b&=&2I^{3L}_b-4s_W^2e_b,\\
a_b&=&2I^{3L}_b,
\end{eqnarray}
where $e_b=-1/3$ is the electric charge 
and $I^{3L}_b=-1/2$ the weak isospin of the $b$ quark.
The new physics contributions $\delta V$ and
$\delta A$ are given by 
\begin{eqnarray}\label{VA}
 \delta V&=&\delta A=\frac{2s_Wc_W}{e}\frac{vm_Z}{\Lambda^2}\left [
   C_{qW}\frac{c_Wm_Z}{2v}-C^{(3)}_{\Phi q} \right ],\\
 S&=&-\frac{8s_Wc_W}{e}\frac{m_b}{\Lambda^2}\frac{v}{\sqrt 2}\left [ 
   C_{Db}\frac{m_Z}{2v}-C_{bW\Phi}c_W\right ].
\end{eqnarray}

In terms of the vertices given in Eq.(\ref{zbb}), 
the observable $R_b$ at LEP I is given by,
to the order of $\frac{1}{\Lambda^2}$, 
\begin{equation}\label{Rb}
R_b=R_b^{SM}\left[ 1+2\frac{v_b\delta V+a_b\delta A}{v_b^2+a_b^2}
      (1-R_b^{SM})\right ],
\end{equation}
where we have neglected the bottom quark mass.
Thus we have
\begin{equation}\label{Vz1}
\delta V=\delta A=
    \frac{R_b^{exp}-R_b^{SM}}{(1-R_b^{SM})R_b^{SM}}
                          \frac{v_b^2+a_b^2}{2(v_b+a_b)}.
\end{equation}
The SM values on $R_b$ and the latest experimental data are [5] 
\begin{eqnarray}\label{data1}
R_b^{SM}&=&0.2158,~~R_b^{exp}=0.2170(9).
\end{eqnarray}
>From Eq.(\ref{Vz1}) and Eq.(\ref{data1}), we obtain at the $1\sigma$ 
($3\sigma$) level 
\begin{equation}\label{bound1}
-0.0053~ (-0.01)<\delta V<-0.0007~ (0.004).
\end{equation}
Assuming that no cancellation between $O_{qW}$ and $O^{(3)}_{\Phi q}$  
occurs, we obtain the bound for each of them at the $1\sigma$ 
($3\sigma$) level
\begin{eqnarray}\label{b1}
-0.5~(-1.03)&<&\frac{C_{qW}}{(\Lambda/{\rm TeV})^2}<-0.07~(0.41),\\
\label{b2}
0.01~(-0.07) &<&\frac{C^{(3)}_{\Phi q}}{(\Lambda/{\rm TeV})^2}<0.09~(0.17).
\end{eqnarray}

Since the operators $O_{bW\Phi}$ and $O_{Db}$ only appear in the $S$ factor
in Eq. (25), their contributions to $R_b$ at LEP I are proportional to 
$m_b/m_Z$ 
and hence suppressed. Therefore, they are not constrained by $R_b$ at LEP I.
However, as Eq. (\ref{wtb}) shows, $O_{bW\Phi}$ induces a right-handed weak 
charged current, and thus it will be constrained by the CLEO measurement on
$b\rightarrow s\gamma$ [9]. The latest limit can be found in Ref. [10], i.e., 
\begin{equation}
-0.03<\frac{C_{bW\Phi}}{\Lambda^2} \frac{\sqrt 2 v m_t}{g_2}<0.00,
\end{equation}
which gives 
\begin{equation}\label{b3}
-0.3<\frac{C_{bW\Phi}}{(\Lambda/{\rm TeV})^2} <0.
\end{equation}

For $O_{t3}$, $O_{Dt}$ and $O_{tW\Phi}$, they are not constrained by $R_b$ 
at the tree level. However, at one-loop level they contribute to gauge
boson self-energies, and thus rather loose bounds exist [7] with significant
uncertainties.  However more reliable, but still rather weak, bounds for 
them can be obtained from the unitarity bound [7]. For $\Lambda=1$ TeV, 
the unitarity bounds are given by [7]
\begin{eqnarray}\label{uni}
\vert C_{t3} \vert <61.5, ~~ \vert C_{Dt} \vert <10.4, ~~
\vert C_{tW\Phi} \vert <13.5.
\end{eqnarray}
\vspace{.5cm}

\begin{center}{\Large 4. Effects on single top quark production at 
                         $e\gamma$ colliders }\end{center}

 Now we examine the effects of the operators in Eqs. (2)-(8) in the single top 
production at high energy $e\gamma$ colliders.  The tree diagrams are
depicted in Fig. 1.  
We note that the effects of $O_{qW}$ is $q^2$-dependent and it can be 
enhanced with higher collider energy.  Since the other operators are 
momentum independent, they do not have the energy enhancement effect.  
Therefore, $O_{qW}$ is of special interest and we will present its 
treatment in some detail.  For the other operators we will only present 
the resultant upper bounds we obtained on their couplings that can be 
probed at the $e\gamma$ collider.
\vspace{.5cm}

\begin{center}{\Large 4.1 Effects of $O_{qW}$}\end{center}

With the effective coupling given in Eqs.(\ref{wtb}) and (\ref{wtbg}), the 
relevant Feynman rules, Feynman diagrams, and Feynman amplitudes are 
implemented in our program package FDC97 [11].  
We take  $m_t$=175 GeV, $m_b$=4.3 GeV, $\alpha_{EW}(m_Z)$=$1/128$,
$|V_{tb}|$=0.9984, $m_Z$=91.187 GeV, $\sin^2\theta_W $=0.232,
$m_W$=$m_Z\cos\theta_W$, $\Gamma_Z$=2.50 GeV and $\Gamma_W$=2.09 GeV.

For $O_{qW}$, the coupling strength 
$C_{qW}/\Lambda^2$ can be determined from Eqs.(\ref{Rb}) and (\ref{VA}),
as a function of $R_b$. In order to show the effect of the constraint from 
the experimental 
value of $R_b$, it will be interesting to investigate the contribution of 
$O_{qW}$ to the 
cross section and the luminosity required to observe the effect
of $O_{qW}$ as a function of $R_b$. 

The center-of-mass energy of the $e\gamma$ is determined by that of the
$e^+e^-$ collider in which one of the lepton beams, say that of the
$e^+$, is back-scattered from a laser beam to produce the high energy 
photon beam.  Hence, the photon beam so produced falls into a spectrum
of energies for a given beam of $e^+$ energy.  We refer to Ref. [12] 
for the details of the photon beam energy spectrum.  The measurable
cross section for the reaction $e\gamma \rightarrow \bar{t}b\nu$ is a
convolution of the reaction cross section at fixed energy with the 
photon energy spectrum.  
As the photon energy spectrum is completely determined 
by the center-of-mass energy of the $e^+e^-$ collider $E_{e^+e^-}$ [12], 
we will use the latter to represent the effective
center-of-mass energy of the $e\gamma$ collider. 
 
In Fig. 2, we show the contribution of $O_{qW}$ to the total cross section 
of the single top production as a function of $R_b$. The solid and dashed 
curves are for $E_{e^+e^-}$  = 500 GeV and 1 TeV, respectively.  
One can see that the experimental data on $R_b$, i.e.,
$R_b^{exp}=0.2170(9)$, has severely constrained
the contributions of $O_{qW}$ to the single top
production rate. For example, with $R_b$ varied within 2$\sigma$, 
i.e., $0.2152<R_b<0.2188$, the contribution of $O_{qW}$ to the
single top rate of the SM prediction is limited to be less
than $8 \%$ and $4 \%$ for $E_{e^+e^-}$ = 1 TeV and 500 GeV, 
respectively.

As we pointed out earlier, since the effects of $O_{qW}$ is
$q^2$-dependent, its effects will be enhanced as the collider 
energy increases. 
In Fig. 3 we plot the total cross sections of single top production, with 
and without the contribution of $O_{qW}$, as a function of center-of-mass
energy. Here we see that the effects of $O_{qW}$ are enhanced significantly
when the center-of-mass energy is above 1 TeV. 

The existence of  $O_{qW}$ will also affect  the 
distribution properties of the final state particles.
In Fig.4, we plotted three distributions,  $d\sigma/d\cos\Theta_{\gamma b}$, 
$d\sigma/dp_t^T$ and $d\sigma/dp_b^T$. Here $\theta_{\gamma b}$ is the angle 
of the $b$-quark with respect to the incident photon direction in the 
$e^+e^-$ rest frame, and $p_t^T$ and $p_b^T$ are 
transverse momentum of  $\bar t$ and $b$ quarks, respectively. 
These figures also show that the new physics contribution
can be enhanced relative to the SM prediction if appropriate cuts 
on the transverse momentum and the angle are imposed. 
We found the optimal cuts to be 
\begin{equation}\label{cuts}
\cos\theta_{\gamma b} \leq 0.84, 
~~p_t^T \geq 70 {\rm GeV}, ~~p_b^T \geq 30 {\rm GeV}.
\end{equation}

To estimate the luminosity required for observing the effects of $O_{qW}$,
we define the significance of the signal relative to
the background in terms of Gaussian statistics, in which a signal at the 
$99\%$ CL is defined by 
\begin{equation}
S \ge 3 \sqrt{S+B},
\end{equation}  
where $S$ and $B$ are the number of signal and background events. 
Applying the cuts of Eq.(\ref{cuts}),  we obtain the luminosity 
required for observing the effects of $O_{qW}$ at 99\% C.L. as
a function of $R_b$.  This is shown in Fig.5,
where the solid and dashed curves are  
for $ 100\% $ and $30 \%$ of detection efficiency, respectively.  
The upper (lower) curves correspond to negative (positive)
values of $C_{qW}/\Lambda^2$.
One can see that if $R_b$ is 
constrained within 2$\sigma$ of the experimental value,
the required integrated luminosity is 4 fb$^{-1}$ (20 fb$^{-1}$)
for $100\%$ ($30\%$) of detection efficiency.
 For a 1 TeV $e\gamma$
collider with $L=10$ fb$^{-1}$ and $30\%$ detection efficiency, the 
absence of a anomalous top quark production will set a bound 
\begin{eqnarray}
-0.85< \frac{C_{qW}}{(\Lambda/{\rm TeV})^2}< 0.38,
\end{eqnarray}
which  is slightly stronger than the corresponding 
$3\sigma$ bound from $R_b$ as given in Eq.(\ref{b1}).
\vspace{.5cm}

\begin{center}{\Large 4.2 Effects of $O^{(3)}_{\Phi q}$, $O_{bW\Phi}$,
$O_{Db}$, $O_{t3}$, $O_{Dt}$ and $O_{tW\Phi}$}\end{center}

The effects of these operators can be analyzed similar to that of 
$O_{qW}$ but they
are momentum independent and thus cannot be enhanced when the collider 
energy increases.  Here we only present the results of our calculation on
the upper bounds of each of them by assuming the existence of one operator 
at a time.  For $E_{e^+ e^-}$ = 1 TeV, $L=10$ fb$^{-1}$ and a  
detection efficiency of $30\%$, they are given by, at 99\% C.L.,
\begin{eqnarray}\label{new1}
-1.9  < \frac{C^{(3)}_{\Phi q}}{(\Lambda/{\rm TeV})^2}&<& 2.1, \\
-0.40 <\frac{C_{bW\Phi}}{(\Lambda/{\rm TeV})^2}&<&0.80, \\
-11.8  < \frac{C_{Db}}{(\Lambda/{\rm TeV})^2}&<& 15.0, \\
-18.5  < \frac{C_{t3}}{(\Lambda/{\rm TeV})^2}&<& 12.8, \\
-4.2  <  \frac{C_{Dt}}{(\Lambda/{\rm TeV})^2}&<& 4.0, \\
-0.65  <  \frac{C_{tW\Phi}}{(\Lambda/{\rm TeV})^2}&<& 0.46 . 
\end{eqnarray}
We note the bounds on $O^{(3)}_{\Phi q}$ and $O_{bW\Phi}$ are also weaker 
than the current ones set by $R_b$ and $b\rightarrow s\gamma$ as given in 
(\ref{b2}) and (\ref{b3}).
For $O_{t3}$, $O_{Dt}$ and $O_{tW\Phi}$, these bounds are much stronger
than their corresponding unitarity bounds which are given by [7] and quoted
in Eq. (\ref{uni}).

\begin{center}{\Large 5. Discussion and summary }\end{center}

In our analyses we assumed the existence of one operator at one time.  
However, if they coexist, their effects have to be disentangled in a 
degree by analyzing additional measurable quantities and by examining 
how all the measurable quantities change under the variation of the 
couplings of the operators involved as shown in Ref. [13] and [14].
Such detailed analyses are not warranted at the present time and we 
will be content ourselves by a few pertinent remarks:

(1) The contribution of  $O_{qW}$ is momentum dependent 
and thus the behavior of the cross section versus $E_{e^+e^-}$ is 
different from
that  predicted by the SM, as shown in Fig. 3. So we can disentangle 
the effects of $O_{qW}$ from  the behavior of the cross section versus 
$E_{e^+e^-}$. 

(2) Since the effects of other operators, 
$O_{\Phi q}$, $O_{bW\Phi}$,  $O_{Db}$, $O_{t3}$, $O_{Dt}$ and $O_{tW\Phi}$,
are momentum independent, the behavior of the cross section with
their contribution versus $\sqrt s$ is the same as that predicted by the SM.
Thus it is hard to distinguish one from another among these operators. 
However, the effects of $O_{\Phi q}$ and $O_{bW\Phi}$, which subject
to similar current constraints as $O_{qW}$, have limited effects at
$e\gamma$ colliders. If the new physics effects at
$e\gamma$ colliders are found to be larger than such limited effects
presented in Figs. 2-4, we could say they may arise from 
$O_{Db}$, $O_{t3}$, $O_{Dt}$ or $O_{tW\Phi}$ and more detailed study 
of these operators are necessary.  
  
To summarize, we used the effective Lagrangian approach to the new physics 
of the top quark to study the possibility of observing anomalous $W tb$
couplings in the single top production at a high energy $e \gamma$
collider. Our results indicate that a luminosity of the order of
100 $fb^{-1}$ is needed 
to reveal the effects of those operators which are subjected 
to the stringent constraints obtained from 
$R_b$ and $b\rightarrow s\gamma$.  
For the operators which are not subjected to any bounds presently, 
meaningful limits of their couplings can be obtained at a $e\gamma $ 
collider. 
\vspace{.5cm}

\begin{center}{\Large  Acknowledgement}\end{center}

We would like to thank E. Boos for discussions.  
J. M. Y. acknowledges JSPS for the Postdoctorial Fellowship.
 This work was supported in 
part by the U.S. Department of Energy, Division of High Energy Physics, 
under Grant No. DE-FG02-94ER40817, by the National Natural Science 
Foundation of China, and by the Grant-in-Aid for JSPS fellows from
the Japan Ministry of Education, Science, Sports and Culture.
\vspace{1cm}

{\LARGE References}
\vspace{0.3in}
\begin{itemize}
\item[{\rm [1]}] R.D. Peccei and X. Zhang, Nucl. Phys. B337, 269 (1990);
                 R.D. Peccei, S. Peris and X. Zhang, Nucl. Phys. B349, 305 (1991);
                 X. Zhang and B.-L. Young,  Phys. Rev. D51, 6564 (1995).
\item[{\rm [2]}] C.T. Hill and S. Parke, Phys. Rev. D49, 4454 (1994);
              D. Atwood, A. Kagan and T. Rizzo, Phys. Rev. D52, 6264 (1995); 
             D.O.Carlson, E.Malkawi and C.-P.Yuan, Phys. Lett. B337, 145 (1994);
             H. Georgi, L. Kaplan, D. Morin and A. Shenk, Phys. Rev. D51, 3888
             (1995);
             T. Han, R. D. Peccei and X. Zhang, Nucl. Phys. B454, 527 (1995);
             K. Cheung, Phys. Rev. D53, 3604 (1996);
             E. Malkawi and T. Tait,  Phys. Rev. D54, 5758 (1996);
             S. Dawson and G. Valencia, Phys. Rev. D53, 1721 (1996);
             T. G. Rizzo,  Phys. Rev. D53, 6218 (1996);
             P. Haberl, O. Nachtman and A. Wilch,  Phys. Rev. D53, 4875 (1996);
             T. Han, K. Whisnant and B.-L. Young and X. Zhang, Phys. Lett. B385, 
                   311 (1996);
             G. J. Gounaris, M. Kuroda and  F.M. Renard, Phys. Rev. D54, 
             6861 (1996);
             G. J. Gounaris, J. Layssac and F. M. Renard, Phys. Rev. D55,  
             5786 (1997);
             G. J. Gounaris, J. Layssac, D.T. Papadamou, G. Tsirigoti 
             and F.M. Renard, hep-ph/9708204;
             G. J. Gounaris, D. T. Papadamou and F. M. Renard, hep-ph/9711399.
             K. Cheung, Phys. Rev. D55, 4430 (1997).
\item[{\rm [3]}] G. Jikia, Nucl. Phys. B374, 83 (1992);
                 E. Boos, Y. Kurihara, Y. Shimizu, M. Sachwitz, H. J. 
                 Schreiber and S. Shichanin, Z. Phys. C70, 255 (1996);
                 E. Boos, A. Pukhov, M. Sachwitz and H. J. Schreiber,
                 Z. Phys. C75, 237 (1997).                  
\item[{\rm [4]}] E. Boos, A. Pukhov, M. Sachwitz, H. J. Schreiber
                   Phys. Lett. B404, 119  (1997).
\item[{\rm [5]}] G. Alteralli,  CERN-TH-97-278, hep-ph/9710434.
\item[{\rm [6]}] C. J. C. Burgess and H. J. Schnitzer, Nucl. Phys. B228, 
                 454 (1983);\\
                 C. N. Leung, S. T. Love and S. Rao, Z. Phys. C31, 433 (1986);
                 W. Buchmuller and D. Wyler, Nucl. Phys. B268, 621 (1986).
\item[{\rm [7]}] G. J. Gounaris, F. M. Renard and C. Verzegnassi, Phys.
                 Rev. D52, 451 (1995);
                 G. J. Gounaris, D. T. Papadamou and F. M. Renard,
                 Z. Phys. C76, 333 (1997);
\item[{\rm [8]}] K. Whisnant, J. M. Yang, B.-L. Young and X. Zhang,
                 Phys. Rev. D56, 467 (1997);   
                 J. M. Yang and B.-L. Young, Phys. Rev. D56, 5907 (1997);
\item[{\rm [9]}] M.Alam et al., CLEO Collaboration, Phys. Rev. Lett. 74, 
                                                          2885 (1995).
\item[{\rm [10]}] M. Hosch, K. Whisnant and B.-L. Young, Phys. Rev. D55, 
                                                        3137 (1997).
\item[{\rm [11]}] J.-X Wang, Computer Phys. Commun. 86 (1993) 214-231;\\
                  J.-X. Wang, preprint KEK-TH-412, 1993,
                    Proceeding of AI93, Oberamergau, Germany;\\
                  J.-X. Wang, Proceeding of AI96, Sept. 1996,
                   Lausane, Switzerland.
\item[{\rm [12]}] K. Cheung, S. Dawson, T. Han, and G. Valencia, Phys.
                  Rev. D51, 5 (1995).
\item[{\rm [13]}] A. Datta, K. Whisnant, Bing-Lin Young and X. Zhang,
                  Phys, Rev. D57, 346 (1997). 
\item[{\rm [14]}] K. Hikasa, K. Whisnant, J. M. Yang and B.-L. Young, 
                  hep-ph/9806401.
\end{itemize}
\eject

\begin{center}Figure Captions\end{center}
\begin{itemize}
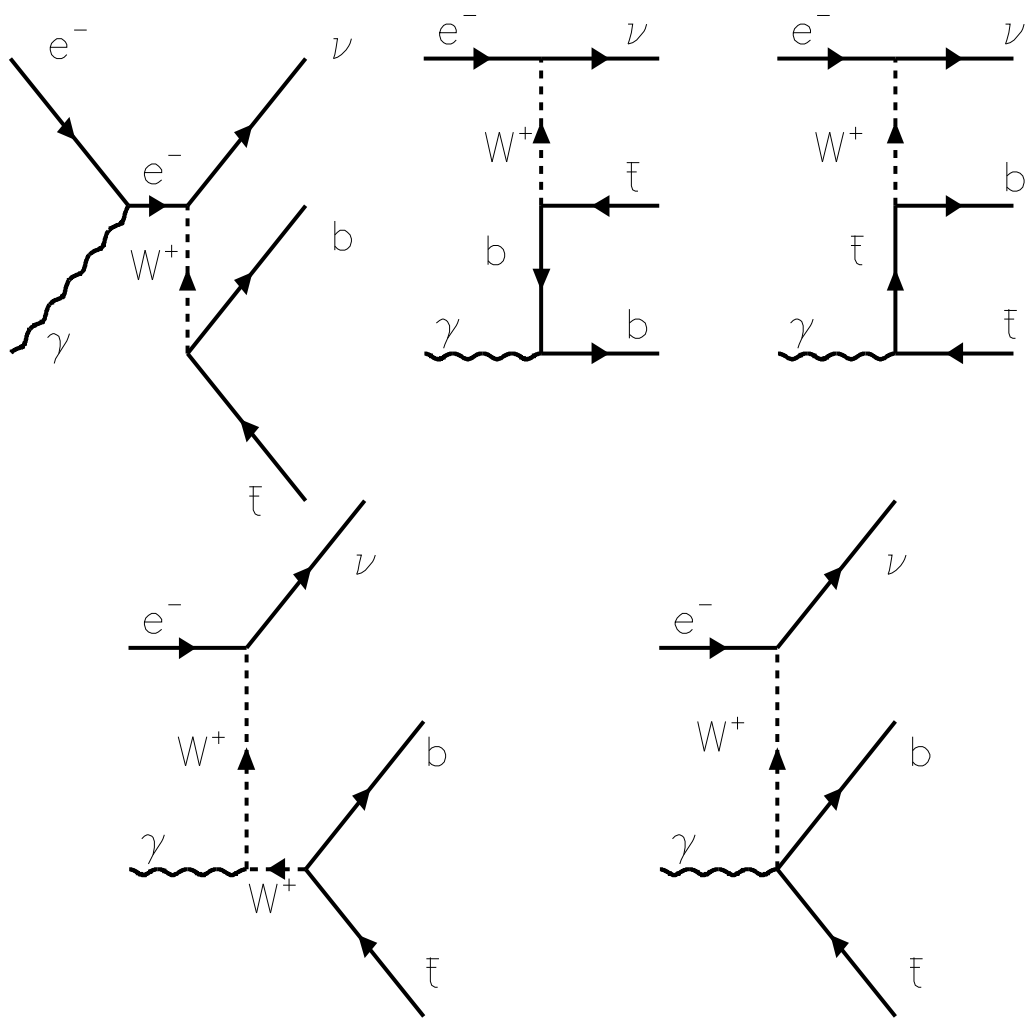
\item[{\rm Fig.1 }] Feynman diagrams for $\gamma e\rightarrow \nu\bar{t} b$.
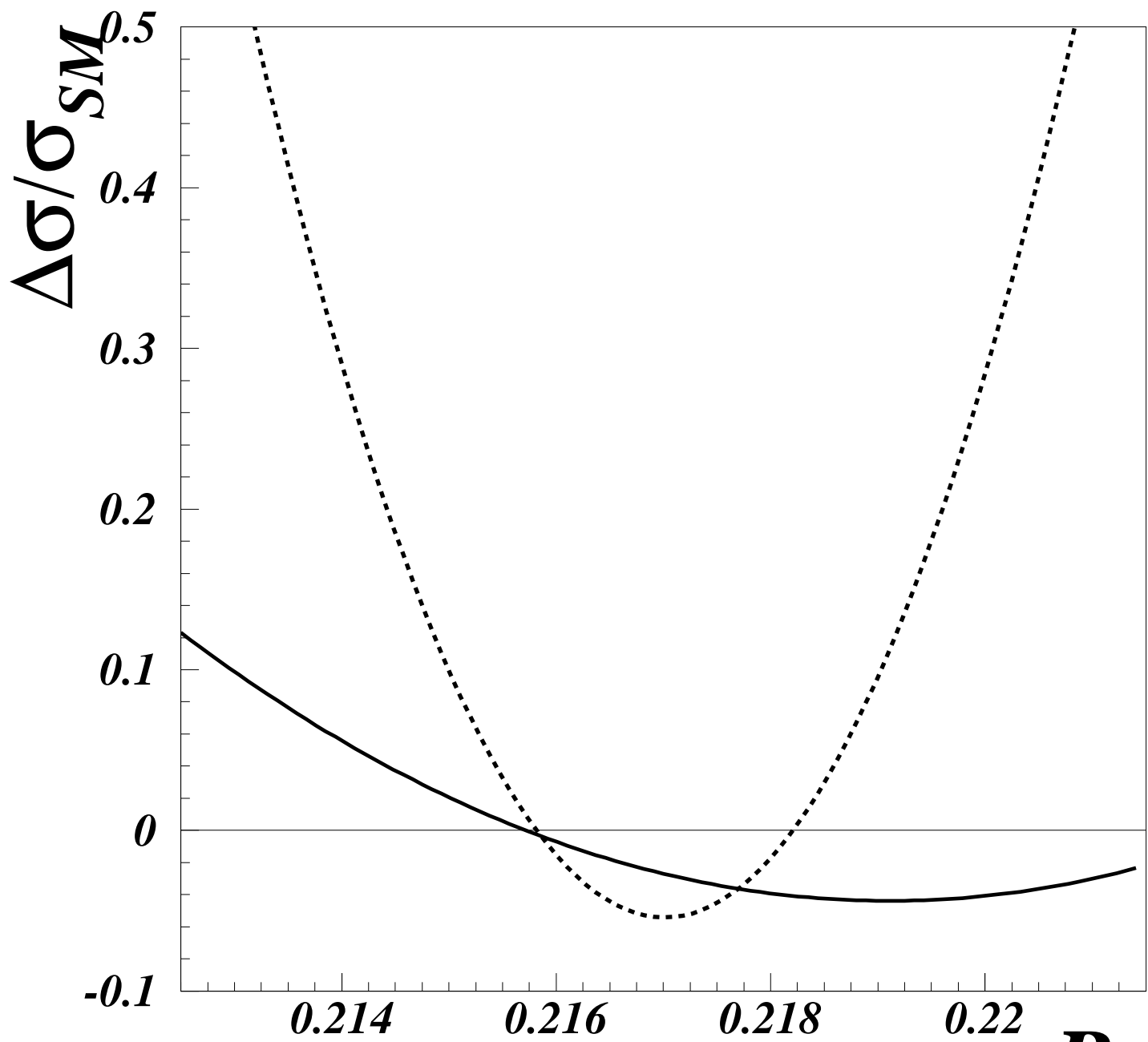
\item[{\rm Fig.2 }] The ratio of $\Delta\sigma/\sigma_{SM}$
                    for the reaction $\gamma e\rightarrow \nu \bar{t} b $
                    versus $R_b$, where 
                    $\Delta\sigma=\sigma-\sigma_{SM}$ with 
                    $\sigma$ and $\sigma_{SM}$ being the cross section 
                    with and without the contribution of $O_{qW}$, 
                    respectively. The solid (dashed) curve is for 
                    $E_{e^+ e^-}$ = 500 (1000) GeV. 
\item[{\rm Fig.3 }] Cross section of reaction 
                    $\gamma e\rightarrow \nu \bar{t}b$
                    as a function of $E_{e^+ e^-}$. The thin line 
                    represents SM expectations, while the dashed and thick 
                    lines are the results with the contribution of $O_{qW}$ 
                    for $R_b = 0.2152 $ and  $R_b = 0.2188$, respectively.
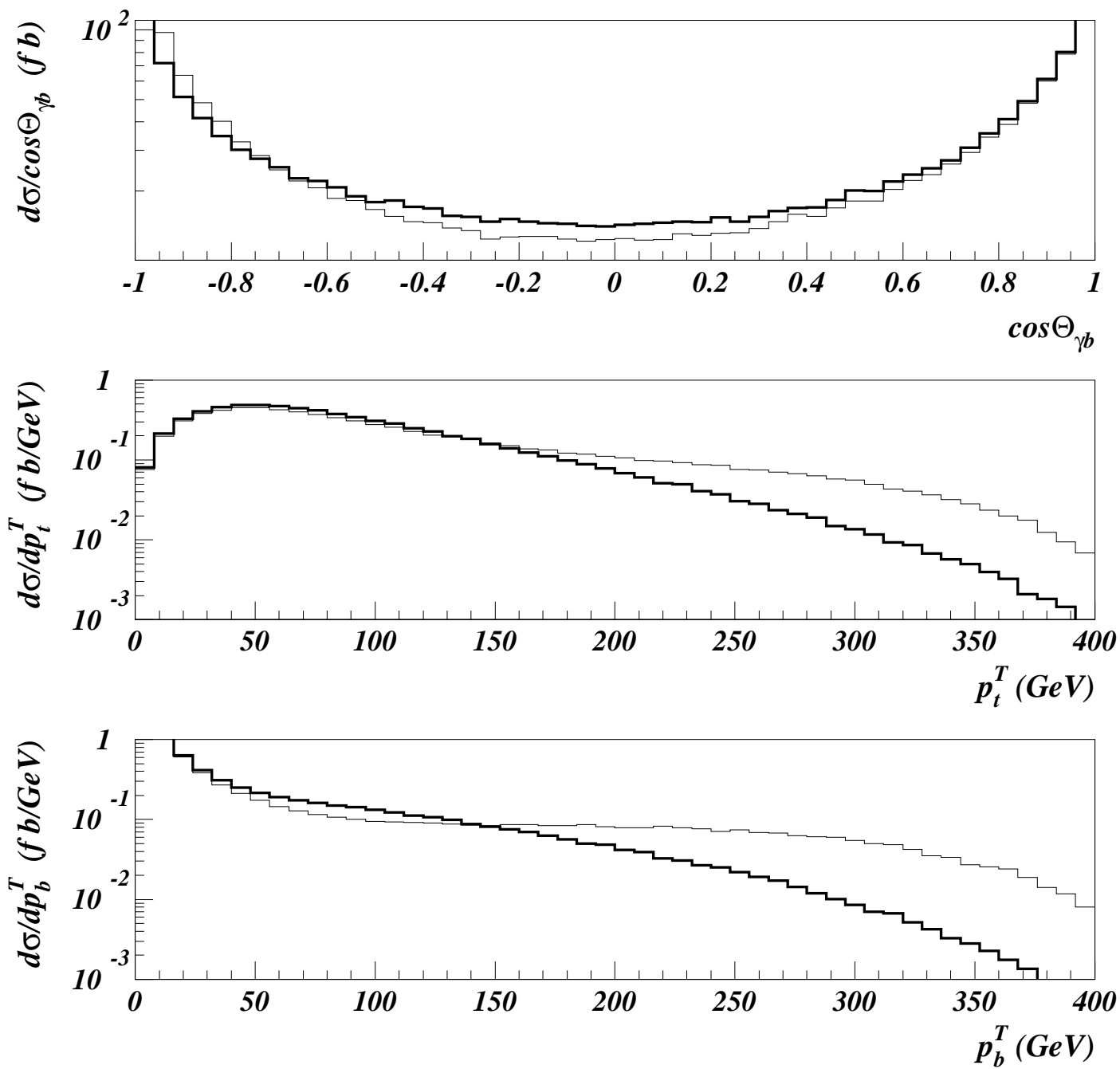
\item[{\rm Fig.4 }] The distribution of differential cross section 
                    of the reaction 
                    $\gamma e\rightarrow \nu\bar{t} b$ versus 
                    $ \cos\theta_{\gamma b}$, $p_t^T$ and $p_b^T$.  
                    The thick and thin lines represent the SM predictions and
                    the predictions with the contribution of $O_{qW}$
                    for $ R_b=0.2188 $, respectively.  
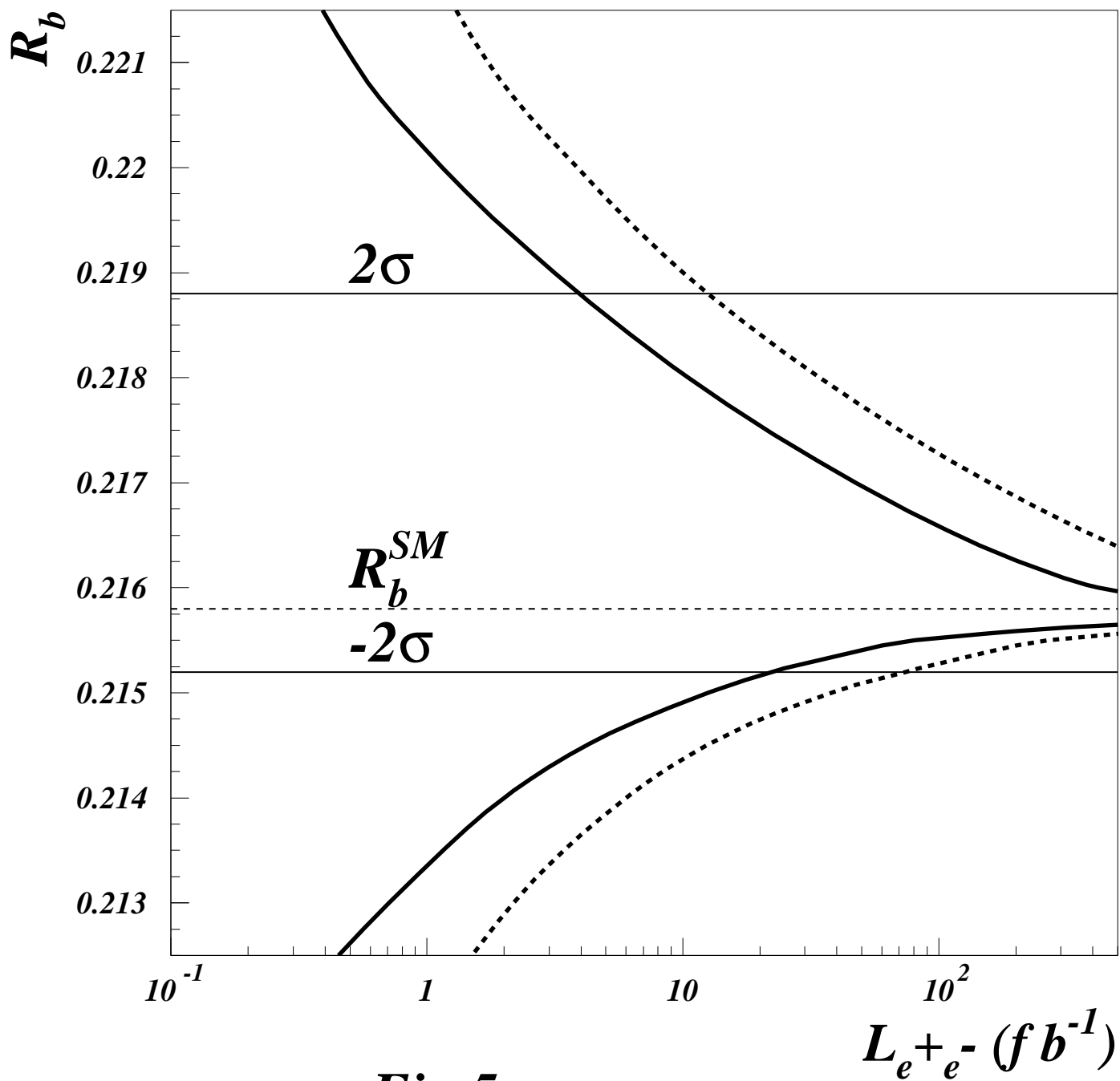
\item[{\rm Fig.5}]  The luminosity required to detect the effects 
                    of $O_{qW}$ at 99\% CL as a function of $R_b$.
                    The solid (dashed) curves are for $ 100\% $
                    ($30 \%$) of event detection efficiency. 
\end{itemize}
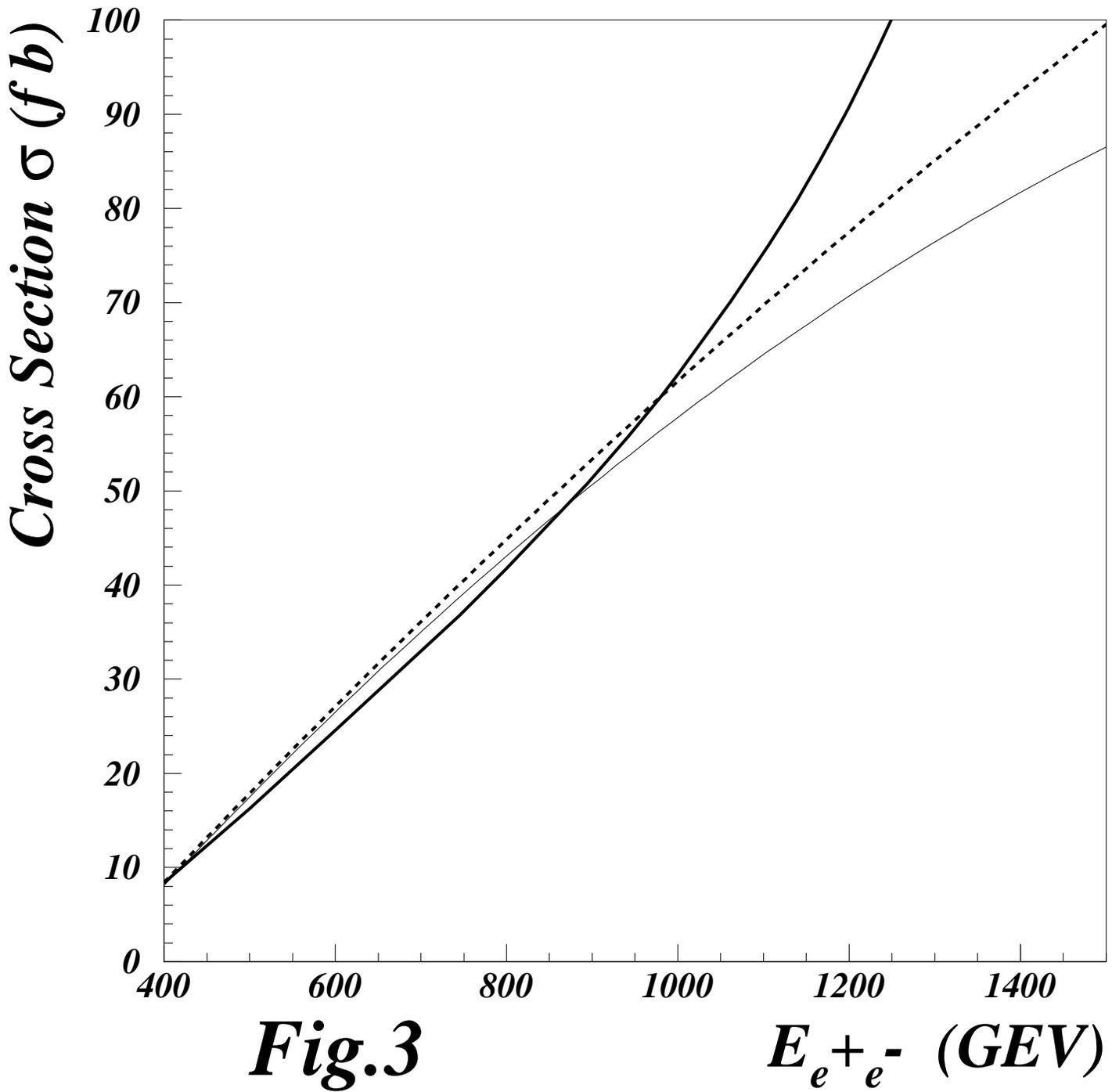
\end{document}